\documentclass[aps,final,twocolumn]{revtex4}%
\usepackage{amsfonts}
\usepackage{amsmath}
\usepackage{amssymb}
\usepackage{graphicx}%
\begin{document}
\title{Magnetic Anisotropy of Isolated Cobalt Nanoplatelets}
\author{T.O.~Strandberg}
\affiliation{Division of Physics, Department of Chemistry and Biomedical Sciences, Kalmar
University, 391 82 Kalmar, Sweden}
\altaffiliation{Division of Solid State Theory, Department of Physics, Lund University, SE-223
62 Lund, Sweden}

\author{C.M.~Canali}
\affiliation{Division of Physics, Department of Chemistry and Biomedical Sciences, Kalmar
University, 391 82 Kalmar, Sweden}
\author{A.H.~MacDonald}
\affiliation{Department of Physics, University of Texas at Austin, Austin TX 78712}
\pacs{PACS number}

\begin{abstract}
Motivated in part by experiments performed by M.H. Pan \textit{et Al.}
\cite{kennanolett05}, we have undertaken a theoretical study of the the
magnetic properties of two-monolayer thick Co nanoplatelets with an
equilateral triangular shape. The analysis is carried out using a microscopic
Slater-Koster tight-binding model with atomic exchange and spin-orbit
interactions designed to realistically capture the salient
magnetic features of large nanoclusters \cite{ac_cmc_ahm2002} containing up to 350
atoms. Two different geometries of the FCC lattice are studied, in
which the nanoplatelet surface is aligned parallel to the FCC (111) and (001)
crystal planes respectively. We find that the higher coordination number in
the (111) truncated crystal is more likely to reproduce the perpendicular easy
direction found in experiment. Qualitatively, the most important parameter
governing the anisotropy of the model is found to be the value of the intra-atomic
exchange integral $J$. If we set the value of $J$ near the bulk value in
order to reproduce the experimentally observed magnitude of the magnetic
moments, we find both quasi-easy-planes and perpendicular easy directions.  At
larger values of $J$ we find that the
easy-axis of magnetization is perpendicular to the surface, and the value of
the magnetic anisotropy energy per atom is larger. The possible
role of hybridization with substrate surface states in the experimental
systems is discussed.

\end{abstract}
\volumeyear{year}
\volumenumber{number}
\issuenumber{number}
\eid{identifier}
\date[Date text]{date}
\maketitle

\qquad

\section{Introduction}
The magnetic properties of transition metal nanostructures are very distinct
from those of bulk materials \cite{AdvPhysmagnano98,skomski_nanomag03}.
Typically, the magnetic moment is enhanced at surfaces \cite{ohnishi83}, in 
ultra-thin magnetic films \cite{Blugel89, Blugel89b},
and in magnetic nanoparticles 
and clusters \cite{billas1994, billas1997, apselPRL96, rusponiNat2003}, 
due to lowered symmetry and
reduced coordination at the surface.
Similarly, one of the most phenomenologically important
properties of magnetic materials,
the magnetic anisotropy energy (MAE) 
is considerably altered in systems of reduced dimensionality.
For example, the magnetic anisotropy per atom in 
two-dimensional (2D) monolayers is more than 
one order of magnitude larger
than in bulk \cite{gambarella02Nat1d}. In  one-dimensional (1D)
magnetic chains the enhancement is 
two to three orders of magnitude \cite{gambarella02Nat1d}. 
It is therefore reasonable to expect
that quasi zero-dimensional nanostructures, that is to say ultra-small nanoparticles
and nanoclusters, should display even more strongly enhanced anisotropy and
other novel phenomena \cite{bansmann05}.

The study of magnetic nanoparticles has a long 
history that goes back to the work of Neel \cite{neel1954} and  
Stoner and Wohlfart \cite{stoner_wohlfart}. 
However, it is only in
recent years that it has been possible to perform experiments on a single magnetic nanoparticle \cite{jamet2001}, 
and to assemble on a metal
surface monolayer magnetic clusters containing up to 
a few tens of atoms \cite{gambarella03Sci}. 
2D monolayer clusters of a few atoms deposited on a metal surface
display a giant magnetic anisotropy perpendicular to the surface
together with unusually large orbital moments \cite{gambarella02Nat1d, gambarella03Sci}
which contribute significantly to the total magnetic
moment of the system, 
unlike in the bulk where they
are almost totally quenched by electron delocalization and crystal field
splitting.  Orbital magnetism and magneto-crystalline anisotropy 
in solids have their common microscopic origin in the spin-orbit
interaction. The enhancement of these two properties in nanoclusters, 
either free or deposited on metallic surface \cite{bansmann05}, 
is a fundamental issue in 
the study of magnetism. Several theoretical calculations based on ab-initio spin-density-functional
theory \cite{nonasPRL2001,lazarovic02} or tight-binding models \cite{pastor1998,
pastor2003} have either predicted 
or partly reproduced these
experimental results. However a number of important points remain unclear:
for example, how the magnetic anisotropy evolves from single-atom to bulk
behavior; how it is affected by the atomic magnetic moments,
how it depends on the arrangement of the atoms in the cluster and 
on the surfaces where the cluster is possibly deposited.

From the point of view of applications,
the remarkable magnetic properties of transition-metal nanoclusters
suggest a potential role as units of ultra-high density magnetic information storage. 
Unfortunately, despite their large anisotropies per atom, these
clusters are still super-paramagnetic at room temperature
\footnote{This is the case also for the system with the largest anisotropy
energy found so far \cite{gambarella03Sci}, namely a single
cobalt atom adsorbed on a platinum surface, which has a MAE of 9 meV,
more than three orders of magnitude larger than in bulk}.
Since the {\it total} anisotropy energy scales roughly with cluster size,
one could in principle try to increase the anisotropy barrier 
and hence the blocking temperature, by increasing
the cluster size. However, experiment \cite{gambarella03Sci} 
has shown that the MAE per atom of clusters
on metal surfaces quickly degrades with the number of atoms.

Recently, the group of M.H. Pan \emph{et Al.} \cite{kennanolett05},
has created arrays of self-assembled Co clusters on top
of a Si (111) surface using epitaxial quantum growth techniques.
In order to prevent silicide formation with the Co, the
surface is covered with half a monolayer (ML) of Al clusters. The resulting
symmetry of the substrate surface prompts the growth of 2 ML thick Co clusters
with equilateral triangle shapes and distinct quantized sizes with sides equal to 
either 2 or 3 times the length of the side of the 7 by 7 Si unit cell - or 5.4 and 8.1
nm, respectively. From the system symmetry and the measured height of the Co
clusters, we propose that the Co atoms position themselves so as to form two
FCC (111) crystal planes, with a modified lattice constant approximately equal
to that of Si.

SQUID magnetometer measurements reveal perpendicular easy directions, and a
very high blocking temperature of 40 $\pm$ 5 K (corresponding to 
an anisotropy energy of 90 meV) and 100 $\pm$ 5
K (corresponding to an anisotropy energy of 220 meV), respectively.
Estimating the number of atoms in the smaller and
larger clusters to be 225 and 484 Co atoms each, this translates to anisotropy
energies of approximately 0.40 and 0.45 meV per atom, corresponding to values
that are one order of magnitude larger than bulk estimates (0.04-0.06 meV/atom
\cite{zangwill2001}). 
The high anisotropy energy of the Co clusters combined with
the Si substrate surface, provides a material that seems ideal for logical circuits
integrated with ultra-high-density memory.

The purpose of this paper is to investigate theoretically the magnetic
properties of nanoclusters similar to those considered 
in the experiment of Ref.~\onlinecite{kennanolett05}.
Using a tight-binding model characterized by short range exchange and atomic 
spin-orbit interactions, we have modeled these clusters to shed light on the 
experimentally observed magnitude of the anisotropy energy and the
perpendicular easy directions.

We find that when the intra-atomic exchange constant $J$ (which is a purely 
phenomenological quantity in our model) is smaller than
1 eV the magnetic moment is in the plane of the nanoplatelet. For larger
values of $J$ - between 1 and 1.5 eV - the easy axis of the magnetization is
is predominately orthogonal to the plane, and the perpendicular anisotropy energy
is in the range of the large values observed in experiment. Associated 
and closely connected with the enhanced magnetic anisotropy, we find that the orbital magnetic moment is also enhanced
and strongly anisotropic, pointing predominantly in the direction of the
easy axis. The stabilization of a perpendicular magnetization orientation
by enhanced anisotropy of the orbital magnetic moment is 
likely related to details of the orbital character of states
near the Fermi level. 

Our motivation for studying the dependence on $J$ is that
the magnetic properties are governed by the ratio of $J$ to the
$d$-electron bandwidth $W_d$. Epitaxial registration of the Co atoms on the Si/Al
surface results in clusters with an effective lattice constant that is larger than
that of bulk Co. This would imply an enhanced value of the ration $J/W_d$ which we try to mimic by
increasing $J$ while keeping $W_d$ constant. In this paper, besides values $J$ between 1.0 and 1.5 eV, which we believe describe realistically our system, we 
also study the asymptotic behavior for the non-physical region $J>1.5$ eV.

This paper is organized as follows.
In section II we outline the theoretical model employed in the analysis
and discuss what can and cannot be learned from its output.
Section III summarizes the
numerical results and the predictions of our model
for the magnetic moment, the anisotropy energy landscape and the anisotropy 
energy per atom as a function of cluster size for two different
geometries of the FCC crystal obtained by truncating it along two different
crystal planes. In Section IV we investigate the
underlying mechanisms responsible for the observed magnetic properties
by studying the dependence of the quasi-particle spectrum as a function of 
the intra-atomic exchange strength $J$. We look at the
change in orbital character and in the distribution of orbital moments of the
quasiparticle eigenstates for increasingly larger values of $J$
Finally, in Section V we summarize our conclusions and suggest an alternative
configuration of substrate atoms that might produce nanoparticles with interesting
magnetic properties. 

\section{Theory}
\subsection{General considerations}
Nanoclusters consisting of ferromagnetic transition metal atoms are
characterized by an unbalanced spin population and a resulting net
magnetization. This paper deals with nanoparticles of sizes below the 10
nm limit, where only a single magnetized domain is observed.
The magnetic anisotropy in small ferromagnetic particles comes from two
distinct sources. Firstly, the long range magnetic dipole interactions cause a
dependence on the overall shape of the particle. Then, there is the short
range exchange interaction, that via the spin-orbit (SO) interaction is
sensitive to all aspects of the electron hopping network, causing a dependence
on crystal symmetry, facet orientation and particle shape. The focus of
our model is this so-called magneto-crystalline anisotropy, which is
responsible for most of the interesting physics in ferromagnetic particles.
The magnetostatic shape anisotropy, due to the magnetic dipole interactions,
can be added as a separate contribution when it is not negligible.

In an infinite crystal, neighbors of a particular site in the lattice will deform the magnetic electron cloud,
which will therefore reflect the point symmetry of the atomic position. The SO
interaction then couples the deformed orbitals to the exchange coupled spins.
In 3$d$ transition metals, the outer, partially filled $d$ shells, are
strongly affected by the crystalline environment. Electronic structure
calculations \cite{jansen1995} imply that most of the anisotropy in this case
comes from a competition between an almost completely isotropic on-site
exchange interaction and strong inter-atomic hopping. The role of the
spin-orbit interaction is that of a relatively small perturbation acting as an
mediator between the two.

The quenching of the orbitals in a crystalline environment breaks the
rotational invariance at each site, and the orbitals are not free to orient
themselves under the influence of an external field. As a result, the atomic
wavefunctions are now no longer eigenstates of $L_{z}$. The new eigenstates
are instead linear combinations for which the expectation value of $L_{z}$
vanishes. In $3d$ transition metals the crystal field splitting is large and
the orbital angular momentum is small. Complete quenching of the wavefunctions
is counteracted by the SO interaction. A microscopic derivation of the magnetic anisotropy
in ferromagnetic materials is highly non-trivial. The most promising attempts
are currently within spin density functional theory.

In a nanoparticle of finite size, the magnetic anisotropy is strongly affected
by the presence of symmetry breaking surfaces.  The surface dependence is sensed 
throughout the system by shape dependent inter-atomic electron hopping paths. In nanoclusters,
the anisotropy is therefore expected to be much larger than in the bulk, and to
depend crucially on the particular shape of the particle. These properties
have been confirmed for Co nanoparticles \cite{jamet2001}.

\subsection{The model}
To construct an effective spin Hamiltonian that describes the exchange
interaction in transition metals, we employ a tight-binding model devised by
A. Cehovin {\it et Al.}\cite{ac_cmc_ahm2002}. 
The aim of this model is to capture generic
features of ferromagnetic metal nanoparticles. Our model Hamiltonian takes the
following form:%
\begin{equation}
\mathcal{H}=\mathcal{H}_{\text{Band}}+\mathcal{H}_{\text{Exc}}+\mathcal{H}%
_{\text{SO}}+\mathcal{H}_{\text{Zeeman}} \label{hmain}%
\end{equation}
Let's address each of the terms in (\ref{hmain}), starting from the left. The
ferromagnetism of transition metals involves itinerant electrons, which
necessitates the use of band theory. The path of the itinerant electrons is
governed by $\mathcal{H}_{\text{Band}}$, describing the orbital motion of the
electrons inside the particle. This is a single-particle tight-binding term,
where the on-site energies of the Wannier orbitals and the hopping matrix
elements between two of them are parameterized by Slater-Koster parameters
\cite{slater_koster}. In neutral Co, we use nine orbitals - one $4s,$ three
$4p$ and five $3d.$ Including the spin degrees of freedom, there are 18
quasiparticle orbitals per atom in our s-p-d tight-binding model.
$\mathcal{H}_{\text{Band}}$ is given in second quantized form by%
\begin{equation}
\mathcal{H}_{\text{Band}}=\sum_{ij}\sum_{s}\sum_{\mu_{1}\mu_{2}}t_{\mu_{1}%
\mu_{2}s}^{ij}c_{i\mu_{1}s}^{\dagger}c_{j\mu_{2}s} \label{ham}%
\end{equation}
where $c^{\dagger}$ and $c$ are the creation and annihilation operators, which
operate on single-particle states labeled by atomic site $\left(  ij\right)
,$ the 9 orbitals $\left(  \mu\right)  $ and spin $\left(  s\right)  $. 
For the $t_{\mu_{1}\mu_{2}s}^{ij}$ we use SK parameters \cite{slater_koster} that
have been extracted from ab initio calculations for the corresponding bulk systems. 

The main purpose of this model is to realistically include the spin-orbit interaction.
The exchange part of the Hamiltonian is simplified using a mean-field
approximation. We first approximate $\mathcal{H}_{\text{Exc}}
$ including only the ferromagnetic exchange, by simply treating the
intra-atomic exchange for $d$-orbitals.%
\begin{equation}
\mathcal{H}_{\text{Exc}}=-2J\sum_{i}\vec{S}_{di}\cdot\vec{S}_{di}%
\end{equation}
where
\begin{equation}
\vec{S}_{di}=\sum_{\mu\in d}\vec{S}_{i\mu}=\tfrac{1}{2}\sum_{\mu\in d}%
\sum_{ss^{\prime}}c_{i\mu s}^{\dagger}\vec{\tau}_{ss^{\prime}}c_{i\mu
s^{\prime}}%
\end{equation}
and $J$ is the parameter that determines the strength of the exchange
interaction. This is set to 1 eV in order to produce an average magnetic
moment per atom of the order of $2\mu_{B},$ somewhat larger than bulk value,
in accordance with calculations \cite{duan2001} and experiment
\cite{billas1994} \cite{billas1997} for Co nanoclusters. $\vec{\tau}$ is a
three-vector with the Pauli matrices as components. We now simplify the
exchange interaction by performing a mean-field decomposition,%
\begin{equation}
\vec{S}_{di}=\langle\vec{S}_{di}\rangle+\delta\vec{S}_{di}%
\label{mf_dec}
\end{equation}
and dropping the second order fluctuation terms in $\delta\vec{S}_{di}.$ The ground state
$\langle\vec{S}_{di}\rangle$ is determined self-consistently. We can now
diagonalize the Hamiltonian numerically, solving a self-consistency condition
iteratively for the mean field order parameters.

Simplifying further still, the exchange mean splitting field is averaged over
all sites, forcing all spins to change their orientation coherently.
\begin{equation}
\vec{h}_{i}\equiv h\hat{\Omega}=\frac{J}{g_{s}\mu_{B}}\langle\vec{S}%
_{di}\rangle\label{mf_ave}%
\end{equation}
This procedure simplifies the anisotropy landscape, neglecting non-collinear
spin configurations. These can occur in nanoparticles, but for
larger clusters most atoms have the same spin
orientations, effectively rendering the system coherent. It is also possible
to prepare a nanoparticle so as to display simple coherent magnetization
reversal processes \cite{wernsdorfer97} \cite{jamet2001}. When inserting the
bulk values for magnetization and spin-splitting field, the mean-field formula
(\ref{mf_ave}) yields the correct value of $J$. This serves as a theoretical
consistency check, but the motivation for the choice of $J$ is that it
produces the experimental and computed mean moment per atom. Selecting a
different $J$ as input parameter, yields a different self-consistent
spin-splitting field. Below we will consider $J$ as a phenomenological 
parameter and we will explore how the magnetic properties of the 
cluster -- in particular, the MAE -- depend on its value.

We expect the mean-field approximation (\ref{mf_dec}), (\ref{mf_ave}) 
to work well, since we are interested only in
particles sizes in the coherent mono domain region. Achieving
self-consistency in the homogeneous average spin splitting field $\vec{h}$,
allows for variations around a mean value in charge density and atomic
moments. With this simplified model, we are able to study larger clusters, up
to sizes of 300-400 atoms, for which the homogeneous exchange field
approximation becomes increasingly accurate.

$\mathcal{H}_{SO}$ is of atomic character, representing the spin-orbit
interaction \cite{bruno89},
\begin{equation}
\mathcal{H}_{SO}=\xi_{d}\sum_{i}\sum_{\mu\mu^{\prime}}\sum_{ss^{\prime}%
}\langle\mu s|\vec{L}\cdot\vec{S}|\mu^{\prime}s^{\prime}\rangle c_{i\mu
s}^{\dagger}c_{i\mu^{\prime}s^{\prime}}%
\end{equation}
where the matrix elements can be explicitly calculated as a function of the
magnetization direction. $\xi_{d}$ characterizes the strength of the SO
coupling, and is taken to be 86 meV \cite{pastor1998}. The spin-orbit coupling
will cause a dependence of total energy on the spontaneous magnetization
direction - the aforementioned magneto-crystalline anisotropy.
The shape of the particle is transmitted to the magnetic anisotropy after many
steps along the electron hopping paths.
The last term in (\ref{hmain}), is a local one-body operator describing the
coupling of the spin and orbital degrees of freedom to an external magnetic
field.
\begin{equation}
\mathcal{H}_{\text{Zeeman}}=-\mu_{B}%
\sum_{i\mu\mu^{\prime}ss^{\prime}}\langle\mu s|\vec{L}+g_{s}\vec{S}|\mu^{\prime}s^{\prime
}\rangle\vec{H}_{\text{ext}}c_{i\mu s}^{\dagger}c_{i\mu^{\prime}s^{\prime}}%
\end{equation}

The simplified mean-field Hamiltonian now appears as follows,
\begin{eqnarray}
\mathcal{H}_{MF}\left(  \vec{h}\right)  & = & \mathcal{H}_{\text{band}%
}+\mathcal{H}_{SO}+\mathcal{H}_{\text{Zeeman}} \\
 & + & \frac{\vec{h}\cdot\vec{h}}{2J}\left(  g_{s}\mu_{B}\right)  ^{2}%
N_{a}-2g_{s}\mu_{B}\vec{h}\sum_{i}S_{di} \nonumber %
\end{eqnarray}
The self-consistent spin-splitting field $\vec{h}^{\ast}$ will minimize the
expectation value of the Hamiltonian, yielding the ground state energy,
$E(\vec{h}^{\ast})=\langle\mathcal{H}_{MF}(\vec{h}^{\ast})\rangle.$ By
diagonalizing $\mathcal{H}_{MF}$ we obtain a set of quasiparticle
eigenenergies, which are occupied up to the Fermi level.

This simplified model of ferromagnetic nanoparticles has the obvious advantage
of being able to treat larger clusters, possessing a much greater
computational simplicity than first principles models. It will provide
us with the generic properties of ferromagnetic nanoclusters, without having to
resort to the much more costly spin density functional theory.
In particular, the detailed properties of the electron hopping network will
depend on atom position relaxations within the nanoparticle.  
We view the model
outlined above having an appropriate level of detail to address properties for
which these relaxations are unimportant. 

In the experiment of Ref.\onlinecite{kennanolett05} 
the distances between two Cobalt atoms,
which are registered to the Si substrate, are larger than bulk Cobalt,
implying that the magnitude of the hopping parameters entering 
the tight-binding model
should be smaller than the bulk values. 
In this paper we did not try to rescale the hopping parameters. 
However, since the crucial parameter that controls the magnetic properties 
of the nanoplatelets is likely
to be $J/W_d$ -- $W_d$ being the width of the $d$-band --
the effect of a smaller $W_d$, could partly be mimicked by
increasing the value of $J$, while keeping the hopping coefficients constant. 
This is one of the reasons of our interest
in studying the $J$ dependence of the magnetic properties.

\subsection{Perturbative analysis}
A qualitative understanding of the system features can be achieved through the
use of perturbation theory. In bulk ferromagnets, the SO interactions are
relatively weak, allowing the use of second-order perturbation theory to
estimate the energy shifts. If we completely turn off the SO interaction, the
eigenstates of a single-particle Hamiltonian are rotationally invariant and of
pure spin character. Turning on the SO interaction, the degeneracy is lifted
and the new eigenfunctions are of mixed spin character. The complete quenching
of the orbital angular momentum is counteracted by the SO interaction that
when treated as a perturbation yields the second-order correction%
\begin{equation}
\varepsilon_{SO}\equiv\varepsilon_{ns}-\varepsilon_{ns}^{\left(  0\right)
}=\frac{\xi_{d}^{2}}{4}\sum_{s^{\prime},m\neq n}\frac{|\langle\psi
_{ms^{\prime}}^{\left(  0\right)  }|\vec{L}|\psi_{ns^{\prime}}^{\left(
0\right)  }\rangle\cdot\vec{\tau}_{ss^{\prime}}|^{2}}{\varepsilon
_{ns}^{\left(  0\right)  }-\varepsilon_{ms^{\prime}}^{\left(  0\right)  }}
\label{pert}%
\end{equation}
where the superscript (0) stands for the unperturbed single particle
wavefunctions and energies, $n$ is an eigenstate label, $s$ a spin label with the 
magnetization orientation direction $\hat{\Omega}$ chosen as the quantization axis,
and $\vec{\tau}$ is a vector containing the three Pauli matrices.
The matrix elements are most easily evaluated by transforming $\vec{\tau}$ to 
the orbital coordinate system. Unlike the infinite solid, where only states of the same $\vec{k}$ that 
are typically separated by an energy of the order of the bandwidth 
$W_{d}$ are coupled, a given state in a nanoparticle is coupled to many other orbitals by the 
spin-orbit interaction.

Simplifying Eq.(\ref{pert}) and averaging over the nine Co orbitals,
we can evaluate the typical SO energy shift as%
\begin{equation}
\varepsilon_{SO}=\frac{\xi_{d}^{2}}{W_{d}}%
\end{equation}
For Co, $\xi_{d}=86$ meV and $W_{d}\sim5$ eV, yielding a typical energy shift
of 1.5 meV. This formula should hold for both bulk material and particles,
provided there are no significant correlations between angular momentum matrix
elements and quasiparticle energy differences.
The anisotropy energy is to a good approximation given by a partially
canceling sum of spin-orbit induced energy shifts depending on magnetic
orientation. Because of these cancellations, the anisotropy energy is in
general much smaller than $\varepsilon_{SO}.$ In a finite system there is
always a perturbative coupling between quasiparticle states close in energy
in Eq. (\ref{pert}), but the matrix elements are distributed among many
states, meaning that typical energy shifts should be comparable to those in
bulk. In general, the anisotropies of particles are larger than those of bulk
because of the loss of symmetry at the surface which tends to reduce cancellations.

\section{Numerical Results}
\subsection{FCC lattice truncation}
We create our nanoplatelets by simply truncating the FCC lattice into 2 ML
thick equilateral triangles, and selecting their surface oriented in parallel
to a given crystal plane. In addition to the experimentally produced clusters,
which we identified as having a surface parallel to the (111) plane, we
also examine clusters that result from choosing a truncation
plane parallel to the (001) crystal surface. These choices yield two different
structures with different symmetries and coordination numbers. The number of
nearest neighbors for an atom in the interior of the particle is 8 for the
(001) geometry, but 9 for the (111).
This difference is reflected in the number of next nearest neighbors, which is
4 for the (001) and 3 for the (111) geometry. The configuration of
next and nearest neighbors will determine the active hopping matrix elements
parameterized by the Slater-Koster parameters in the term $\mathcal{H}_{\text{Band}}$
in the Hamiltonian (\ref{ham}), i.e. the orbital motion of the electrons
inside the nanoparticle.

\begin{figure}[ptb]
\includegraphics[width=8cm]{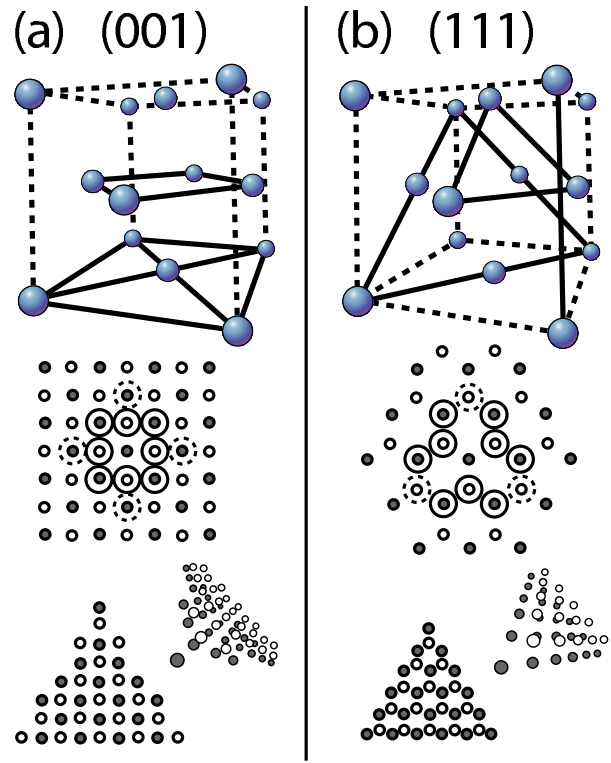}\caption{
The structure
resulting from choosing two different planes of truncation for the 2 MLs of
the nanoparticle. The left column (a) shows the effect of choosing the
nanoplatelet surface parallel to the (001) FCC crystal plane and the right
column (b) depicts the situation obtained by instead selecting the (111) FCC
crystal plane. The top row shows the FCC unit cell with the nanoparticle atoms
connected by solid lines. In second row there is a top view of the 2 MLs
selected in the unit cells above. The dashed circles indicate the next nearest
neighbors solid circles indicate the nearest neighbors of an atom in the interior of the nanoparticle.
Finally, the bottom row shows the resulting nanoparticle (top view and
in perspective).}%
\label{truncations}%
\end{figure}

\begin{figure}[ptb]
\includegraphics[width=4cm]{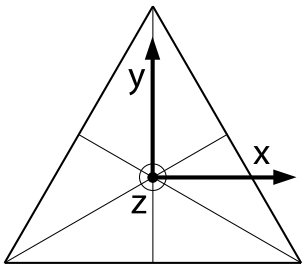} \caption{We define our
coordinate axis relative to the triangular nanoplatelet, with the z-axis
perpendicular to the surface, and the y-axis parallel to the surface and
in a mirror symmetry plane (unique in the case of (001)).}%
\label{axis}%
\end{figure}

Examples of structures resulting from choosing two different truncation planes are shown
in Fig.~\ref{truncations}. We note that the resulting (111) geometry has a
higher degree of symmetry with a three-fold rotation axis running
perpendicularly through the triangle surface and three mirror planes, whereas
the (001) version only has a mirror symmetry across the vertical
axis. These symmetries must all be preserved in the magnetic anisotropy energy
landscape resulting from the particle. One additional symmetry will always be
present in all anisotropy landscapes, irrespective of shape or size. Due to
time-reversal invariance we will always have the inversion symmetry,%
\begin{equation}
E\left(  \theta,\varphi\right)  =E\left(  \pi-\theta,\pi+\varphi\right)
\end{equation}
where $\theta$ and $\varphi$ are the spherical coordinates that define the
magnetization direction, $\hat{\Omega}.$ 
Throughout this article, we will fix our coordinate system as in
Fig. \ref{axis}, with the z-axis $(\theta=0)$ perpendicular to the nanoplatelet surface
and the y-axis $(\theta=\pi/2,\varphi=\pi/2)$ parallel to the surface and in one of the mirror symmetry planes 
of the cluster.
In addition to the above described clusters have also examined one monolayer
thick clusters, obtained by simply removing the top layer of a (111) cluster
of a given size.

\subsection{Magnetic Moments}
\begin{figure}[ptb]
\includegraphics[width=7.5cm]{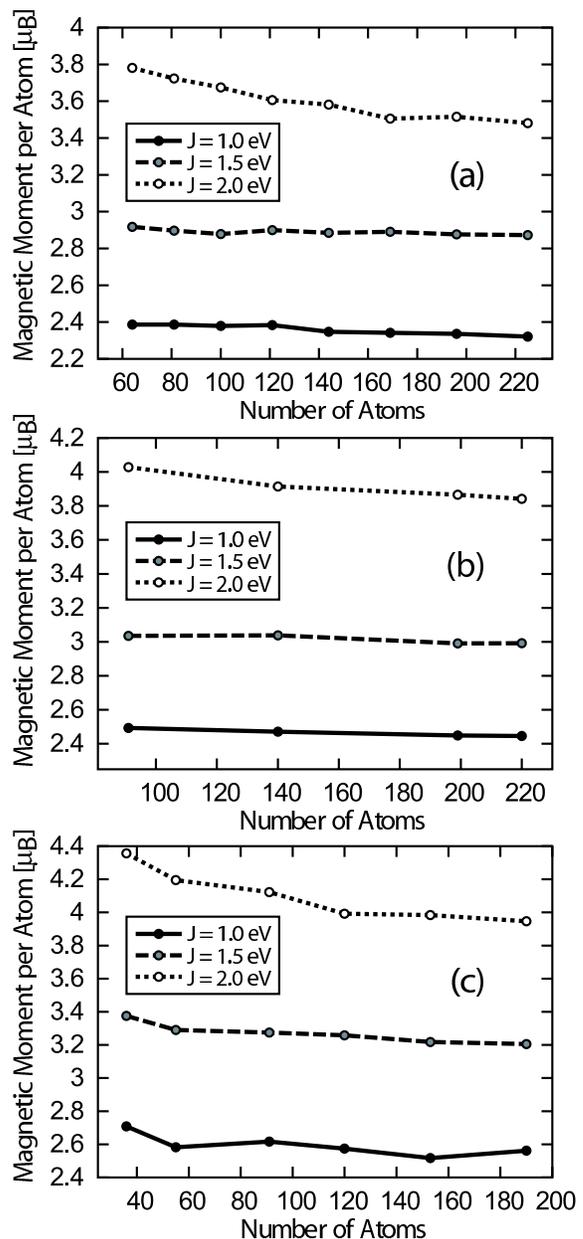}\caption{The atomic mean magnetic
moment as a function of cluster size for the different $J$ and all geometries -
(a) 2 ML (111), (b) 2 ML (001) and (c) 1 ML (111). The relatively
elevated values of 1 ML (111) is associated with imbalance effects brought on
by a larger surface area. The experimental mean magnetic moment
is approximately $2.1\pm0.2\mu_{B}$ \cite{kennanolett05}$,$ which is similar to the value obtained
for 2 ML clusters and $J$ = 1.0 eV.}%
\label{moments}%
\end{figure}

Fig.~\ref{moments}
shows the mean magnetic moment per atom for increasing cluster size with one curve for
$J=1.0,1.5$ and $2.0$ eV. The increase in the mean
moment with $J$ occurs because the system more strongly favors the
alignment of spins in the $d$-channel, and the spin-character of the
eigenstates becomes more well defined. This behavior is associated with a
redistribution of $d$-charge into the $p$- and $s$-channels. 
By altering the electronic configuration of the $3d$ levels and
increasing the number of singly occupied levels, the total energy is
decreased. We will address this issue in more detail below.

The atomic mean magnetic moment curves level out above 100 atoms and we can extract
the approximate values for the smallest of the experimentally produced
nanoplatelets, i.e. the one consisting of roughly 225 atoms. The mean magnetic
moments per atom are presented in table \ref{momtab}. The 1 ML value corresponds to
taking the top layer off the 225 atom 2 ML (111) version, yielding a cluster
consisting of 120 atoms.

\begin{table}[ptb]%
\begin{tabular}
[c]{|r|c|c|c|}\hline
& 2 ML (111) & 2 ML (001) & 1 ML (111)\\\hline
$J=1.0$ eV & 2.32$\mu_{B}$ & 2.44$\mu_{B}$ & 2.57$\mu_{B}$\\\hline
$J=1.5$ eV & 2.87$\mu_{B}$ & 2.99$\mu_{B}$ & 3.99$\mu_{B}$\\\hline
$J=2.0$ eV & 3.48$\mu_{B}$ & 3.84$\mu_{B}$ & 4.31$\mu_{B}$\\\hline
\end{tabular}
\caption{The mean magnetic moment per atom for the $\sim$225 atom clusters (the 1 ML
cluster has 120 atoms)}%
\label{momtab}%
\end{table}

We can compare these values with the experimentally observed value for the mean
magnetic moment $\mu=2.1\pm0.2\mu_{B}$ \cite{kennanolett05}. Only the $J=1.0$ eV
setting will produce a similar value.  We nevertheless include larger values of 
$J$ in the following calculations in order to address the physics underlying 
anisotropy energy trends. 
Both 2 ML clusters display similar behavior, but the 1 ML (111) cluster
exhibits a slight relative enhancement of the mean magnetic moment. In the 2
ML clusters, a second monolayer induces cancellation effects, with a
subsequent drop in the mean moment value. In general, the number of unbalanced
spins increases with the surface to volume ratio of the nanoparticle. 
The difference between the 2 ML (001) clusters and the 2 ML (111) clusters
can be attributed to the lower symmetry of the (001) cluster.

\subsection{Magnetic Anisotropy Landscapes}
Magnetic anisotropy energy landscapes have been computed for the three geometries
described in the previous subsection (1 and 2 ML (111) and 2 ML (001)) for
increasingly larger cluster sizes, up to approximately 225 atoms.

The results of these calculations reveal two major types of anisotropy
landscapes. First, we have the Quasi-Easy Plane (QEP) which consists of a
discrete number of in-plane energy minima, corresponding to easy directions
consistent with the symmetries of the particle. These minima are separated by
very low energy barriers, thus forming a quasi-easy plane oriented in parallel
with the surface of the nanoplatelets. The second type of observed anisotropy
landscape, corresponds to a bistable system with two easy directions oriented
perpendicularly to the cluster surface, separated by a large blocking barrier.

We have explored the parameter space governing the resulting anisotropy
landscape keeping track of three variables: the type of geometry,
the size of the cluster ($\lesssim225$ atoms), and the value of the exchange
coupling strength ($J=1.0$, 1.5 and 2.0 eV). Qualitatively, we can separate our results into
two size regimes - a small cluster regime consisting of 50 atoms and less, and
a large cluster regime consisting of clusters that have more than 50 atoms. 
\begin{figure}[ptb]
\includegraphics[width=8.5cm]{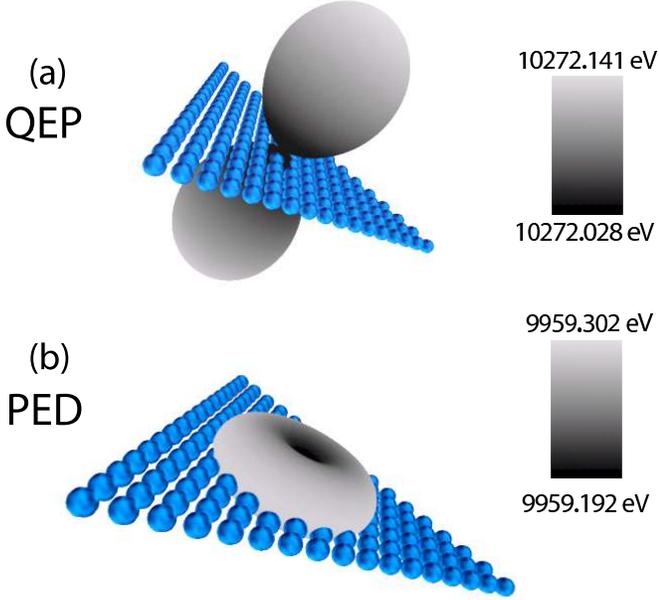}
\caption{The two typically 
observed anisotropy energy landscapes exemplified by 153 atoms of 1 ML (111). (a) shows the Quasi-Easy
Plane (QEP) solution and (b) the landscape with Perpendicular
Easy Directions (PED). These graphs show how the energy varies with the
direction of the self-consistent spin splitting field on the unit sphere, and
have been rescaled so that the minimum energy is located at the origin and the
maximum at unity (see Eq. (\ref{rescx})-(\ref{rescz})). For clarity, the energy value has
also been color coded and the individual atoms superimposed. The colorbars on the right show the grayscale interpolation
between the maximum (top, white) and minimum (bottom, black) energy in eV.} 
\label{landscapes}%
\end{figure}

In the large cluster regime ($>50$ atoms) all geometries exhibit
quasi-easy planes for $J=1.0$ eV. In contrast, for $J=1.5$ and $2.0$ eV all of
these clusters produce landscapes with Perpendicular Easy Directions (PEDs).
The two typical landscapes always take an approximate form exemplified in
Fig.~\ref{landscapes} by 153 atoms of 1 ML (111). In this figure, the energy
landscape on the unit sphere has been renormalized according to the equations 
\begin{eqnarray}
x  & = &\frac{E\left(  \theta,\varphi\right)  -E_{\min}}{E_{\max}-E_{\min}}%
\sin\theta\cos\varphi \label{rescx} \\
y  & = &\frac{E\left(  \theta,\varphi\right)  -E_{\min}}{E_{\max}-E_{\min}}%
\sin\theta\sin\varphi\\
z  & = &\frac{E\left(  \theta,\varphi\right)  -E_{\min}}{E_{\max}-E_{\min}}%
\cos\theta
\label{rescz}
\end{eqnarray}
so that at any given point the distance from the origin corresponds to the
energy value, with the minimum value located at the origin and a maximum value
at a distance of unity. $\theta$ and $\varphi$ refer to the polar and
azimuthal angle, respectively. In addition to the rescaling, the distance from
the origin has been color coded for viewing convenience - the black minima
that appear in these shapes correspond to the easy directions.

Turning to the more sensitive regime of small clusters ($\leq$ 50 atoms), the
1 ML (111) and the 2 ML (001) clusters display only QEPs for $J=1.0$ eV. For
the same setting of $J$ and 2 MLs of (111), we instead find a wide range of
intermediate shapes between the QEP and PED landscapes. Most notably, the 2 ML
(111) clusters are the only ones capable of producing a clear bistable system
with PEDs for the original setting of $J=1.0$ eV. This behavior indicates that
the 2 ML clusters generated in the (111) plane are more prone to forming a
perpendicular bistable system than their (001) counterparts. The qualitative
behavior of the anisotropy landscapes in the two different size regimes 
is summarized in table \ref{qeped}. Note QEP/PED means that
quasi-easy planes (QEP), perpendicular easy directions (PED) and intermediate
shapes between these are observed in the small cluster regime.
No qualitative change in any of the anisotropy landscapes can be
observed for $J>2.0$ eV, indicating that we have already reached a saturation
of the effect brought on by increasing the exchange.

\begin{table}[ptb]%
\begin{tabular}
[c]{|c|c|c|c|}\hline
$\mathcal{N}_a \leq 50$ & 2 ML (111) & 2 ML (001) & 1 ML (111)\\\hline
$J=1.0$ eV & QEP/PED & QEP & QEP\\\hline
$J=1.5$ eV & QEP/PED & QEP & PED\\\hline
$J=2.0$ eV & PED & PED & PED\\\hline\hline
$\mathcal{N}_a>50$ & 2 ML (111) & 2 ML (001) & 1 ML (111)\\\hline
$J=1.0$ eV & QEP & QEP & QEP\\\hline
$J=1.5$ eV & PED & PED & PED\\\hline
$J=2.0$ eV & PED & PED & PED\\\hline
\end{tabular}
\caption{Qualitative behavior of the clusters with less than (top)
and more than (bottom) 50 atoms.}%
\label{qeped}%
\end{table}

\begin{figure}[ptb]
\includegraphics[width=6.5cm]{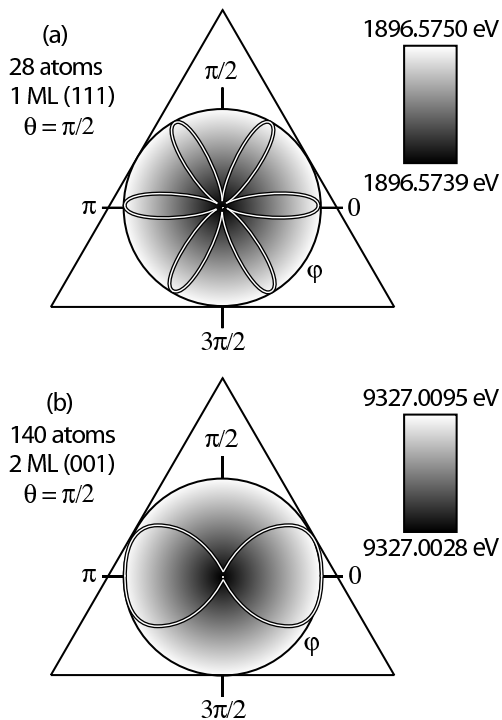}\caption{Examples of the
discrete minima structure in the quasi-easy planes for (a)
1 and 2 ML (111) and (b) 2 ML (001). The black and white lines in the circle trace out the section
of the MAE landscape in the xy-plane parallel with the cluster surface $(\theta=\pi)$,
renormalized so that the origin corresponds to the energy minimum and the radius
of the circle to the (local) maximum (see Eq. (\ref{rescx})-(\ref{rescz})). The grayscale background
in the circle shows how the energy varies with distance from the origin, 
with minimum energy at the origin (black) and maximum at the circle radius (white).
The numerical values of the max.~and min.~are given at the top and bottom of the
colorbar on the right. The six- and two-fold minima are consistent with
the symmetries of the particle. Note that the (111) in-plane barrier is significantly
smaller than that of (001).}%
\label{minima}%
\end{figure}

Traces of the symmetries of the particle resulting from the chosen geometry
are present in the associated anisotropy landscapes with a varying
degree of visibility.  The way in which the symmetry of the particle is reflected in the
anisotropy landscape is most easily identified by considering the
configuration of discrete minima in the quasi-easy plane landscapes (see
Fig.~\ref{minima}). These figures show the local maxima/minima structure in the xy-plane
that coincides with the surface of the nanoplatelets. 

Group theoretical considerations for the FCC lattice
geometry predict a six minima structure. Although we break the full FCC
symmetry by selecting only a few monolayers, this six minima structure is
still present in the (111) clusters, identifiable as the easy directions
parallel to the symmetry lines of the equilateral triangle running from the
corner to the midpoint of the opposing line. In the (001) case the symmetry is
even more badly broken and there are only two remaining minima corresponding
to in-plane easy directions parallel to the mirror symmetry line, running
vertically through the center of the particle. Similar symmetry effects can be
seen to a lesser extent in the perpendicular easy direction landscapes, either
as a hint of a hexagonal shape replacing the circular rim of the torus in the
case of the (111), or as a weakly oval toroid in the (001) case.

\begin{table}[ptb]%
\begin{tabular}
[c]{|r|c|c|c|}\hline
& 2 ML (111) & 2 ML (001) & 1 ML (111)\\\hline
$J=1.0$ eV & -0.10 meV & -0.69 meV & -1.02 meV\\\hline
$J=1.5$ eV & 0.58 meV & 0.49 meV & 0.74 meV\\\hline
$J=2.0$ eV & 0.02 meV & 0.17 meV & 0.45 meV\\\hline
\end{tabular}
\caption{The anisotropy energies per atom corresponding to the $\sim$225 atom
nanoplatelets.}%
\label{ergtab}%
\end{table}

\subsection{Anisotropy Energies}
\begin{figure}[ptb]
\includegraphics[width=7.3cm]{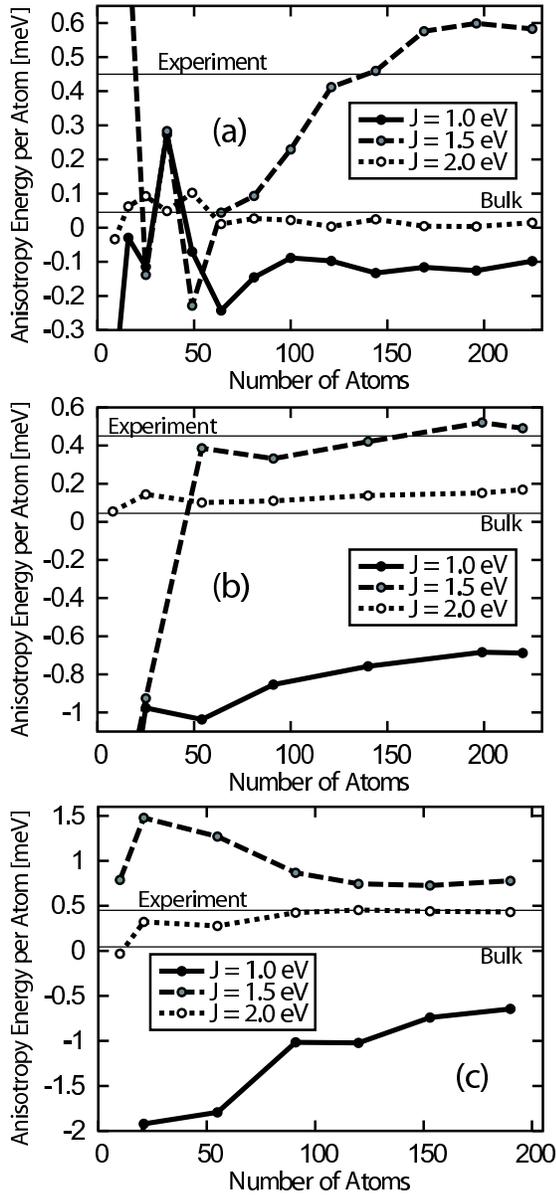}\caption{The anisotropy energy
per atom as a function of cluster size for different J and the three
different geometries - (a) 2 ML (111) , (b) 2 ML (001) and (c) 1 ML
(111). Each curve represents a different value of the exchange coupling. A negative 
sign indicates QEPs and a positive PEDs. Some
fluctuations are to be expected, particularly for the smallest clusters, where
the mean-field treatment is not so accurate. 1 ML (111) displays a generally higher
anisotropy energy due to larger unbalanced spin populations.} 
\label{energies}%
\end{figure}

Fig.~\ref{energies} shows how the anisotropy energy per atom varies with the
number of atoms. As before, each diagram represents a different geometry, 
and each curve represents a specific value of $J$. The anisotropy energy
is defined with a sign as the in-plane direction energy minus the perpendicular direction
energy, so that a negative value corresponds to QEPs and a positive to PEDs.
The horizontal lines
mark the experimental ($\approx0.4$ meV) and bulk ($\approx0.045$ meV) value.
Table \ref{ergtab} displays the anisotropy energy per atom for the approximate
sizes corresponding to the smaller of the two dimensionally quantized clusters
grown (225 atoms).

We expect fluctuations in the anisotropy energy to be large for the smaller
clusters, where our mean-field treatment is less accurate. It is obvious
however, that the fluctuations are larger for the (111) in comparison to
(001). 
Some fluctuations can be caused by the failure of mean field
theory, but in the 2 ML (111) case the fluctuations persist even for relatively
large clusters between 50-100 atoms. The fluctuations in this region for
the 2 ML (111) $J = 1.5$ eV curve, can be attributed 
to the phenomenon of weakly avoided level-crossings \cite{ac_cmc_ahm2002}, that can occur
as the model parameter space is varied - particularly for smaller clusters.

\begin{figure}[ptb]
\includegraphics[width=6.5cm]{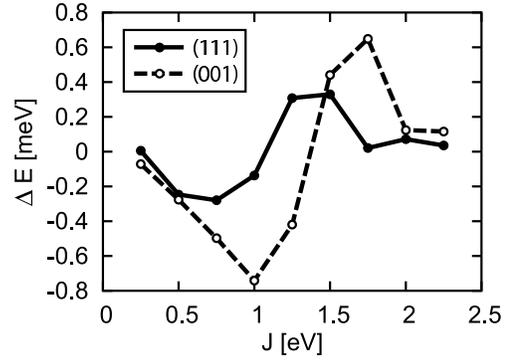}
\caption{The anisotropy energy as a function of $J$ for $\sim$ 140 atom
clusters of 2 ML (001) and (111) reveals a curve that
roughly agrees with what one might expect from perturbation theory. 
The transition from QEP to PED (minus to plus) occurs for smaller $J$ in the 
case of the (111) cluster, indicating a bias toward PED for this geometry.}%
\label{jdep}%
\end{figure}

We note that the ordering of the absolute magnitude of the energies with respect to $J,$ is not the
same for the different geometries. For 2 ML (001) and 1 ML (111) the anisotropy energy magnitude
follows a generally decreasing trend with increasing $J$. In the case of 2 ML (111)
the $J=1.5$ eV curve starts to climb rather steeply and overtakes the $J=1.0$ eV magnitude
around 100 atoms. For a given cluster size, perturbation theory predicts that the magnitude of the anisotropy
energy will initially increase as a function of small $J$, and decrease asymptotically
for large $J$. The turning point of the anisotropy magnitude as a function of the exchange strength will
be different for the different geometries, which explains the different ordering of the magnitudes 
in figure \ref{energies}. In figure \ref{jdep} we have mapped out the exchange strength versus anisotropy energy
in order clarify this difference between the (001) and the (111) geometries.

\section{Dependence of quasiparticle spectrum on exchange coupling}
In order to shed light on the overall
magnetic behavior of the nanoplatelets described above,
in this section we study the properties of the underlying quasiparticles 
and how they depend on the
exchange coupling $J$.

\subsection{Quasi-particle energy levels and eigenstates}
\begin{figure}[ptb]
\includegraphics[width=6cm]{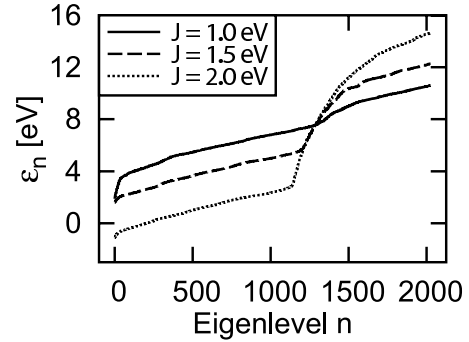}
\caption{The eigenenergies of
all occupied levels for 225 atoms of 2 ML (111). The spin-aligned
region (below the curve discontinuity) of pure $d$-character lowers the total
energy via the exchange term. This is at the cost of a rise in the total energy due to the higher,
mixed states. The resulting total energy is minimized by balancing the energy
cost and gain associated with these two regions. All geometries display a similar linear low
energy region and a discontinuous mid-spectrum leap.}%
\label{eigen}%
\end{figure}

\begin{figure*}[ptb]
\includegraphics[width=16cm]{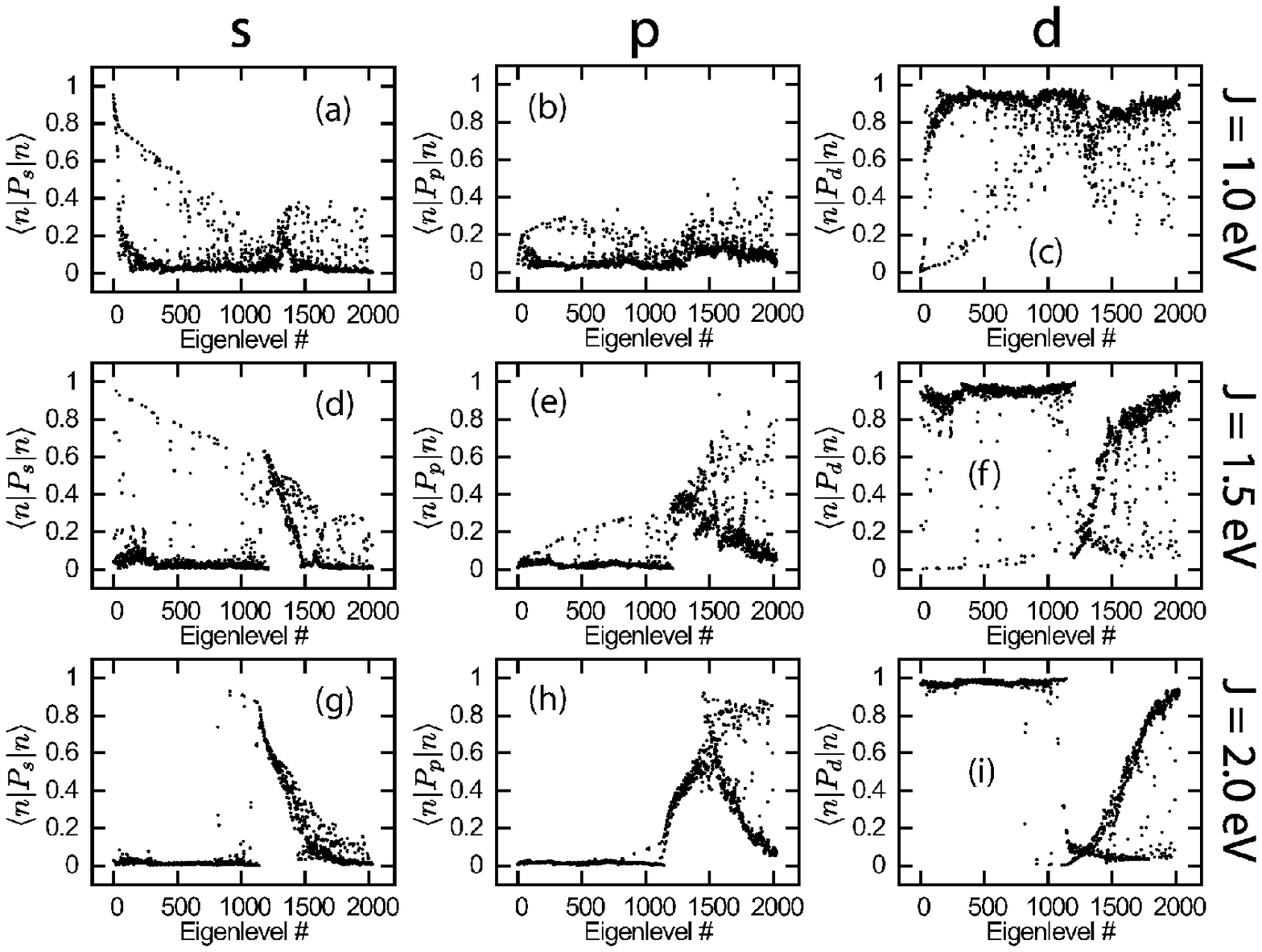}\caption{The orbital
character of the occupied eigenlevels for a 225 atom 2 ML (111) cluster. Each
column shows the respective $\ell$-channel obtained by applying the associated
projection operator as defined in Eq. (\ref{proj}), and each row refers to a
different value of the exchange coupling. Increasing $J$ prompts the formation
of a virtually pure $d$-character region covering the lower half of the 
eigenspectrum, and the levels with high $s$- and $p$-level mixing are focused 
on the upper half of the spectrum. Note that the high $d$-mixing returns just below the Fermi
surface. The other geometries display a qualitatively similar evolution of the orbital character mixing 
pattern with increasing $J$.}%
\label{spd144}%
\end{figure*}

\begin{figure*}[ptb]
\includegraphics[width=16cm]{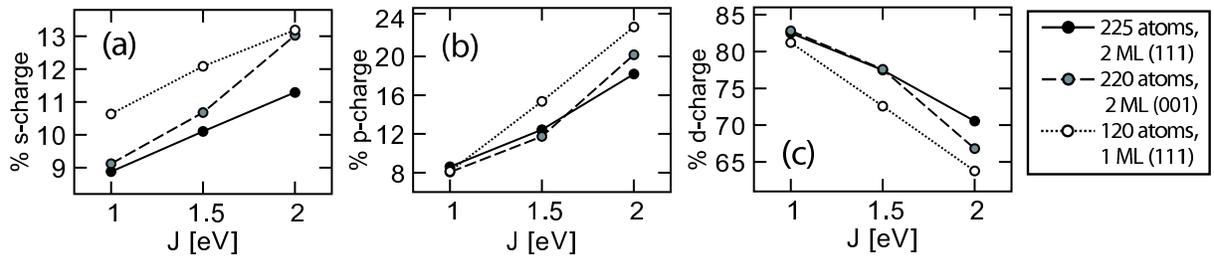}\caption{The total
percentage of $\ell$-charge for the representative clusters. Orbital
character changes such that the $s$- and $p$-channels become more populated at
the expense of the $d$-channel with increasing $J$.}%
\label{percent}%
\end{figure*}
We start by looking at
the ``band structure'' for three different values of $J$. 
In Fig.~\ref{eigen} the quasiparticle
energies are plotted as a function of eigenvalue index up to Fermi level for 225 atoms of 2 ML (111).
The behavior for the three
systems considered so far-- 2 ML (111), 2 ML (001) and 1 ML (111)--
is qualitatively similar. 
The spectrum is characterized by a low-energy part that increases linearly
with the eigenvalue index and extends up to roughly mid spectrum where there
is a shoulder that becomes more pronounced with increasing $J$. 
As we show below, the low-energy states are majority-spin states 
of predominant $d$-character. From the figure we can see that increasing
$J$ simply causes a rigid down-shift of this part of the spectrum, while
the width of the spectrum about the shoulder increases considerably.

In order to determine the orbital mixing of the eigenlevels, we can project
out the $s,p$ and $d$ characters using the associated projection operator.%
\begin{equation}
P_{\ell}=\sum_{i}\sum_{s}\sum_{\mu_{\ell}}|i,\mu_{\ell},s\rangle\langle
i,\mu_{\ell},s| \label{proj}
\end{equation}
where $\ell=s,p,d$ and the projection operators fulfill the sum rule
\begin{equation}
\sum P_{\ell}=P_{s}+P_{p}+P_{d}=1
\end{equation}
The result of this operation is shown in figure \ref{spd144} for 225 atoms of
2 ML (111), where the three orbital channels have been separated. At $J=1.0$
eV, the mixing is fairly homogeneous over the range of eigenlevels, with an
overshadowing total percentage of $d$-character, as is expected for Co.
Remarkably, as $J$ is increased, redistribution of charge results in a
configuration where roughly all levels in the lower half of the spectrum obtain a pure $d$-character.
Above this eigenlevel, two distinct lines of mixing can be observed. The first
one skims the bottom of the $d$-channel and represents levels with a very low
$d$-character mixing. Secondly, there is a line of mixing representing the
other half of the eigenlevels in the upper region, telling us that they obtain
an increasing mixture of $d$ with the eigenenergy, such that at the Fermi
surface, the character is almost 100\% $d$ again.

Increasing $J$, the $s$-charge vacates the lower region, making way for the $d$-charge and
lodges mostly in a narrow region above the pure $d$-levels. 
The $p$-charge exhibits a similar evolution pattern, but ends up with a maximum in
the center of the upper half of the spectrum. The other geometries - 1 ML (111)
and 2 ML (001) - display a qualitatively similar behavior when projecting out the different channels.

Fig.~\ref{percent} shows the total percentages of $\ell$-charge in the
nanoparticle. By inspection, we see that the above described segregation of
orbital character of the eigenlevels with the energy, is associated with a
general funneling of $d$-charge mainly into the more energetic $p$-levels, but
we also find a smaller percentage going into the $s$-channel. Comparing the
2 ML (001) and 2 ML (111) curves, we see that the
redistribution when $J$ is increased is more pronounced for the (001) symmetry.

\subsection{Quasi-particle spin}
\begin{figure*}[ptb]
\includegraphics{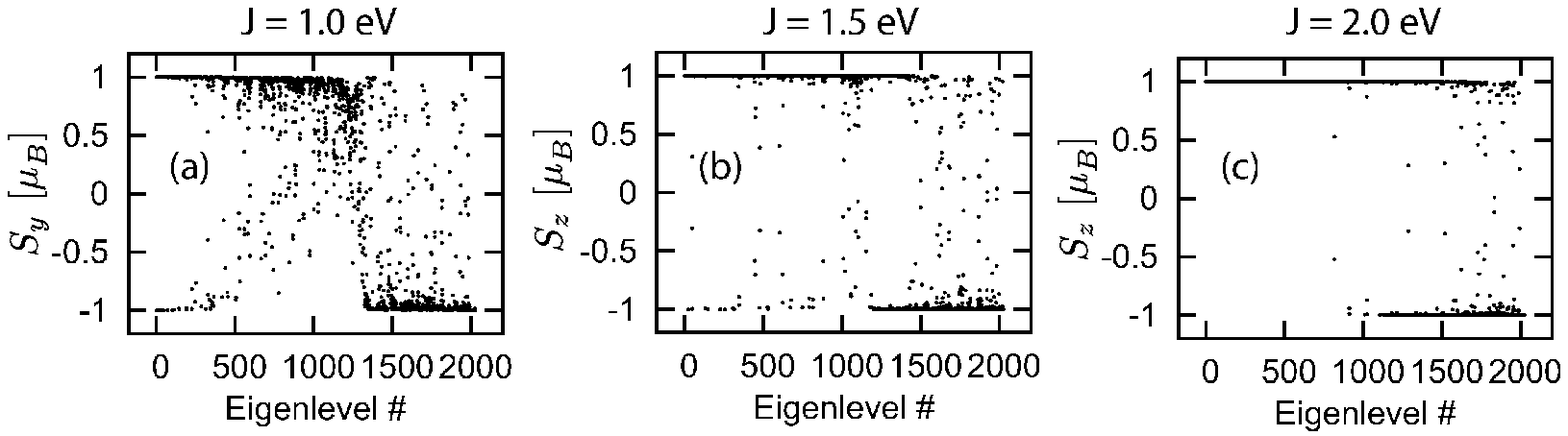}\caption{The single-particle 
expectation value of the spin component along the easy axis (the direction
of the nanoparticle magnetization).
The spin character of the quasi-particle states 
becomes more well defined with increasing $J$.
These results are for a 225 atom  2 ML (111) nanoplatelet.
The other
geometries display very similar evolution patterns with increasing $J.$}%
\label{spins}%
\end{figure*}
We next discuss the $J$-dependence of the 
spin character of the quasi-particle states.
It is useful to define our axis relative to
the triangular nanoplatelet as in figure \ref{axis}, with the $z$-axis
perpendicular to the surface, and the $y$-axis parallel to the nanoplatelet
surface and one of the triangles symmetry lines.
In Fig.~\ref{spins} we plot the quasi-particle expectation values
$\langle n|S_i|n\rangle$
of the component of the spin operator $S_i$ in the direction of 
the magnetization, that is, along the $y$-axis for for $J=1.0$ eV and
along the $z$-axis for $J=1.5$ eV and 2.0 eV. We have verified that the
expectation value of the component of the spin operator in other directions
is negligible. The numerical results shown here are for the 225 atom 2 ML (111) system.
The other two geometries considered display a very similar trend.
As anticipated in the previous section, when we discussed the energy spectrum
in Fig.~\ref{eigen}, for $J=1.0$ eV the low-energy states up
to the energy shoulder correspond to (predominant)
majority-spin states; above the shoulder, at the top of the majority-spin band,
the minority-spin
band starts. For $J=1.0$ eV there are however several states 
whose spin is not well-defined.
When $J$ increases to 1.5 and 2.0 eV, most of the states acquire a well-defined
spin character and there is a conversion of 
minority-spin states into majority spins leading to an increase of the
net spin magnetic moment (see Fig.~\ref{moments}). As we can see 
in Fig.~\ref{spd144}, the
change in spin character is also accompanied by a change in orbital character,
in which the $d$-states are redistributed into the $p$- and $s$-channels.  
Nevertheless the Fermi levels still lies 
among states of predominant $d$-character.
For $J=1.5$ and $2.0$ eV the majority- and minority-spin bands overlap. 
The shoulder seen in Fig.~\ref{eigen} occurs at the bottom of the minority-spin
band.
For $J=2.0$ eV  the majority-spin density of states at the Fermi level is
not zero but 
much smaller than the minority-spin density of states.


\subsection{Quasi-particle anisotropy energies}
The quasi-particle anisotropies are obtained by taking the difference between
the eigenenergies of level $i$ in the hard and easy direction.
Table \ref{astddev} displays the standard deviations of the distributions for the three
different clusters. 
Typically, the width of the distribution is between one to two
order of magnitude larger than the mean value, indicating that the resulting
sum of anisotropies has large cancellations and that individual levels may shift the
value of the total anisotropy. 

\begin{figure}
\includegraphics[width=6.5cm]{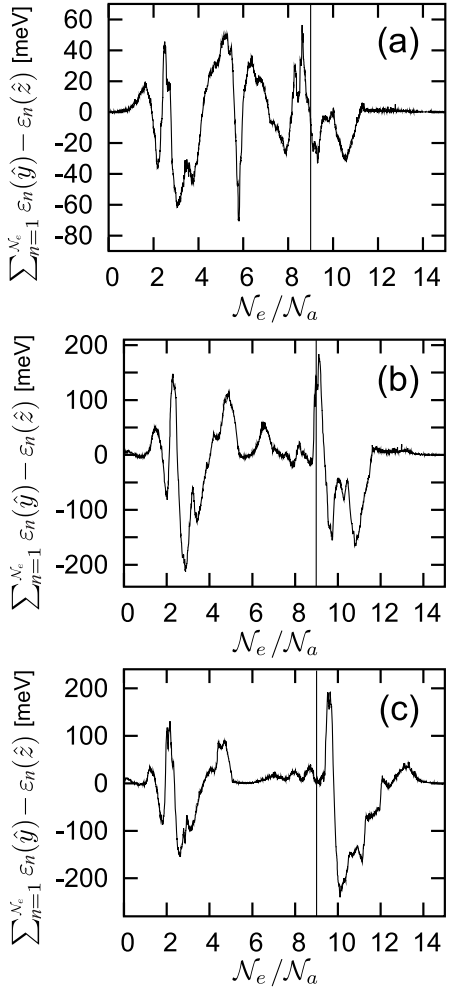}
\caption{Plot of the cumulative sum of single quasiparticle anisotropies (see Eq. \ref{cumuleq}) for 225 atoms of 2 ML (111) as a 
function of the ``band filling''
${\mathcal{N}_e}/{\mathcal{N}_{\rm A}}$, where ${\mathcal{N}_e}$ is the total
number of occupied quasiparticle states and ${\mathcal{N}_{\rm A}}$
is the number of atoms.
The vertical lines mark the band filling for cobalt. The three panels refer
to: (a) $J= 1.0$ eV, (b) $J=1.5$ eV, and (c) $J= 2.0$ eV respectively.}
\label{cumulative}
\end{figure}

In Fig.~\ref{cumulative} we plot the cumulative sum of 
individual quasi-particle anisotropy energies, i.e.
\begin{equation}
{\mathcal{C}}({\mathcal{N}_e})= \sum_{n=1}^{\mathcal{N}_e}\hspace{1mm}
[ \varepsilon_{n}(\hat{y})-\varepsilon_{n}(\hat{z})]\;,
\label{cumuleq}
\end{equation}
where we sum over an increasing total number of occupied quasiparticle  
states ${\mathcal{N}_e}$, 
and display the result as a function of the ``band filling'', that is,
${\mathcal{N}_e}$ divided by the
number of atoms ${\mathcal{N}_{\rm A}}$. The vertical line
at ${\mathcal{N}_e}/{\mathcal{N}_{\rm A}}=9$ corresponds to the cobalt band
filling.
The result reveals large and rapidly varying 
fluctuations of ${\mathcal{C}}({\mathcal{N}_e})$ 
as a function of the band filling.
This supports the idea that the total anisotropy energy is given to a 
good approximation by
a partially canceling sum of quasiparticle anisotropies,
and that only a fraction of states below the Fermi 
level is responsible for the overall anisotropy-energy characteristics
of the system \cite{ac_cmc_ahm2002}. 
For the particular case of the 2 ML (111) shown in the figure, we can see that  
the cumulative density at cobalt band filling as a function of $J$
experiences a sharp maximum in the 
vicinity of $J=1.5$ eV, which accounts for the behavior of the total
anisotropy energy displayed in 
Fig.~\ref{energies}.
Similar sharp patterns
are present in the cumulative sum for clusters 
resulting from the other geometries.

\begin{figure}[ptb]
\includegraphics[width=6cm]{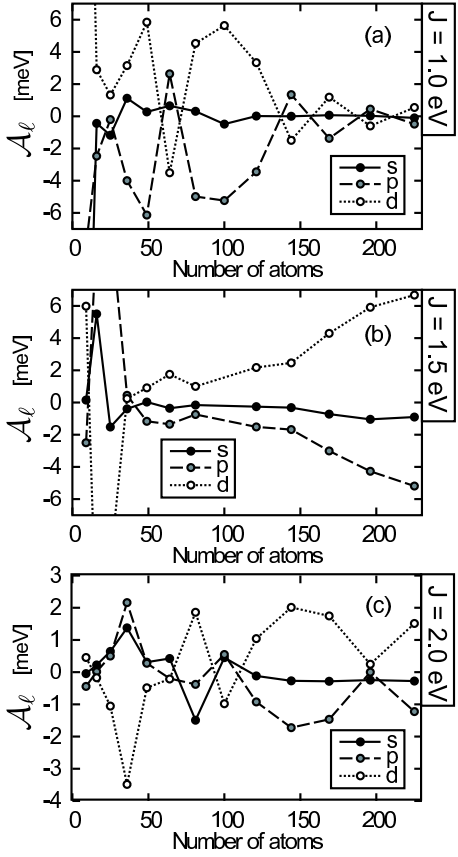} %
\caption{The contribution to the anisotropy from each $\ell$-channel as a function of cluster size for the 2 ML (111) geometry.}
\label{esep}
\end{figure}

Fig.~\ref{esep} shows the contribution to the total anisotropy energy (with sign)
coming from each $\ell$-channel for the 2 ML geometries.
This orbital character filtering of the single-quasiparticle levels
has been calculated using the equation
\begin{eqnarray}
\nonumber
\mathcal{A}_{\ell} & \equiv &
\sum_{n=1}^{N} \hspace{2mm}[  \hspace{1mm} \langle n\left(  \hat{x}\right)  |P_{\ell}|n\left(
\hat{x}\right)  \rangle\varepsilon_{n}\left(  \hat{x}\right)  \\
&  & \hspace{10mm} - \langle n\left(  \hat{z}\right)  |P_{\ell}|n\left(  \hat{z}\right)  \rangle
\varepsilon_{n}\left(  \hat{z}\right)  \hspace{1mm}]
\label{projanis}
\end{eqnarray}
where $\varepsilon_{n}(\hat{y})$ \& $\varepsilon_{n}(\hat{z})$ are the single particle 
eigenenergies for eigenstates $|n(\hat{y})\rangle$ \& $|n(\hat{z})\rangle$,
corresponding to the solutions with the field constant $\vec{h}$ oriented in
parallel to the $y$ and $z$-axis respectively (see figure \ref{axis}).
In this way, we obtain a measure
of the total contribution to the anisotropy energy coming from the electrons of orbital
character $\ell$.

From Fig.~\ref{esep} we can see that in general the major contribution to 
the anisotropy energy comes from states of $d$-character, which tend to
favor perpendicular anisotropy (note however that for small
clusters, $d$-states can have in plane anisotropy). States of $s$
character contribute little except for very small clusters.
States of $p$-character mostly favor in-plane
anisotropy and their relative contribution can be non-negligible.
This trend is consistent with the fact that, as shown in Fig.~\ref{spd144},  
a few eigenstates around the Fermi level have strong $p$ character when
$J \ge  1.5$ eV, but the anisotropy is still dominated by the most
numerous $d$-character states.

\begin{table}[ptb]%
\begin{tabular}
[c]{|c|c|c|c|}\hline
\emph{System} & $J$ [eV] & \emph{Type} & $\sigma$ [meV]\\\hline\hline
225 atoms of & 1.0 & QEP & 1.31\\\cline{2-4}%
2 ML (111) & 1.5 & PED & 2.47\\\cline{2-4}
& 2.0 & PED & 2.07\\\hline\hline
220 atoms of & 1.0 & QEP & 2.61\\\cline{2-4}%
2 ML (001) & 1.5 & PED & 1.93\\\cline{2-4}
& 2.0 & PED & 1.70\\\hline\hline
120 atoms of & 1.0 & QEP & 5.35\\\cline{2-4}%
1 ML (111) & 1.5 & PED & 4.52\\\cline{2-4}
& 2.0 & PED & 4.11\\\hline
\end{tabular}
\caption{Standard deviations for the distributions of
single-level anisotropies 
as a function of increasing $J$.}%
\label{astddev}
\end{table}

\subsection{Orbital contribution to the magnetic moment}

Finally, we consider the $J$-dependence of the quasi-particle orbital
moments, defined as the expectation value of the orbital angular momentum 
$L$ with respect to the quasi-particle states $|n\rangle$. 
The matrix elements
contain the normal local part of the orbital angular momentum, but also the
non-local part originating from delocalized electrons under the influence
of the spin-splitting field \cite{ccm2003_gfpap7}.

In the absence of SO interactions the orbital moment is quenched.
When SO is present the orbital moments are non-zero and are believed
to play a crucial role in the magnetic properties of magnetic nanostructures.
For our magnetic nanoplatelets we find that: (i) the orbital moment 
is larger than the bulk value and contributes significantly 
to the total magnetic moment of the system; (ii) the orbital moment is
strongly anisotropic and its direction
essentially coincides with the direction of the spin moment; (iii) its 
magnitude decreases as a function of $J$.

We are interested in the distribution of quasiparticle orbital contributions 
to the magnetic moment coming from
all occupied eigenlevels computed when the spin-splitting field is chosen
in the direction
of the self-consistent solution. 
We compute expectation values in these states 
of the out-of-plane component $L_z$
and the in-plane-component $L_y$. Note that for $J=1.0$ eV, 
when the magnetization is in the
$xy$-plane, the $y$-axis corresponds to one of the discrete minima for
the total energy (see Figs.~\ref{minima} and \ref{axis}) and we choose this
direction for the self-consistent field.

\begin{table}[ptb]%

\begin{tabular}
[c]{cc||c|c|c|c|}\hline
\multicolumn{1}{|c}{\textit{System}} & \multicolumn{1}{|c||}{$J$ [eV]} &
$\overline{\langle L_{z}\rangle}$  & $\overline{\langle L_{y}\rangle}$ &
$\sigma_{z}$ & $\sigma_{y}$ \\\hline\hline
\multicolumn{1}{|c}{225 atoms of} & \multicolumn{1}{|c||}{1.0} & -0.007 &
0.20 & 0.76 & 0.11\\\cline{2-6}%
\multicolumn{1}{|c}{2 ML $\left(  111\right)  $} & \multicolumn{1}{|c||}{1.5}
& 0.18 & -0.006 & 1.53 & 0.05\\\cline{2-6}%
\multicolumn{1}{|c}{} & \multicolumn{1}{|c||}{2.0} & 0.08 & -0.002 & 1.94 &
0.06\\\hline\hline
\multicolumn{1}{|c}{220 atoms of} & \multicolumn{1}{|c||}{1.0} & 0.003 &
0.25 & 0.24 & 0.13\\\cline{2-6}%
\multicolumn{1}{|c}{2 ML $\left(  001\right)  $} & \multicolumn{1}{|c||}{1.5}
& 0.26 & 0.001 & 0.29 & 0.05\\\cline{2-6}%
\multicolumn{1}{|c}{} & \multicolumn{1}{|c||}{2.0} & 0.09 &
$<10^{-5}$ & 0.28 & 0.05\\\hline\hline
\multicolumn{1}{|c}{120 atoms of} & \multicolumn{1}{|c||}{1.0} & -0.007 &
0.27 & 0.98 & 0.16\\\cline{2-6}%

\multicolumn{1}{|c}{1 ML $\left(  111\right)  $} & \multicolumn{1}{|c||}{1.5}
& 0.09 & -0.003 & 2.15 & 0.07\\\cline{2-6}%
\multicolumn{1}{|c}{} & \multicolumn{1}{|c||}{2.0} & 0.04 & -0.001 & 2.78 &
0.08\\\hline
\end{tabular}

\caption{Statistics for the orbital matrix elements in units of $\mu_{B}$. 
Column 3 \& 4 display the atomic mean orbital moment and the last two columns
the standard deviation for the distribution of all eigenlevels. Large contributions to the matrix elements
come only from the components along the easy-axis. }%
\label{lstat}%
\end{table}

For the quasi-in-plane $J=1.0$ eV solution, we find that there are
sizable distributions in both $y$ and $z$ directions.  
Note that the $x$-component (not shown here) is negligibly small ($\leq
10^{-10}$). Looking closely at the features of these distributions by means
of Tab.\ref{lstat}, we can see that the width of $z$-component is seven times
wider than the $y$-component. However, the $z$-component is very symmetric
around zero. Therefore the resulting average value is small 
$\overline{\langle  L_y\rangle} \approx 0.007 \mu_B$.
In contrast, the distribution of the $y$-component is skewed around zero
and its average is not small, 
$\overline{\langle  L_z\rangle} = 0.20 \mu_B$. 
This value corresponds to an orbital moment $= 0.20 \mu_B$, 
larger than the cobalt bulk orbital moment $0.14\mu_B$ per atom\cite{chen95}.
As $J$ is increased to 1.5 eV, the magnetization direction switches from
in-plane to out-of-plane. The distribution of the $z$-component widens
whereas the $y$-component narrows. Furthermore, it is now the average orbital
moment in the $z$-direction that is the dominant one 
($\overline{\langle  L_z\rangle} = 0.18 \mu_B >> \overline{\langle  L_y\rangle} = 0.006 \mu_B$), 
although slightly smaller than
the $y$-contribution of the $J=1.0$ eV case.
When $J$ is further increased to 2 eV, both 
$\overline{\langle  L_z\rangle}$ and 
$\overline{\langle  L_y\rangle} = 0.006 \mu_B$ decrease.
and become even smaller than the bulk value.
The behavior for the the other two types of clusters is 
qualitatively similar.

Since the
the orbital
magnetic moment is strongly anisotropic and it is large
when the total magnetic anisotropy is large, we can try to connect 
these two quantities. We follow Bruno's perturbative argument
and we write \cite{bruno89}
\begin{equation}
E_{\rm ani} = E(\hat z) - E(\hat y) = - \frac{1}{2}\frac{\xi(J)}{2 \mu_B} 
(m_{\rm L}^z - m_{\rm L}^y)\;,
\label{brunoeq}
\end{equation}
where $m_{\rm L}^{z,y} = \mu_B\overline{ \langle L_{z,y}\rangle}$.
For  ferromagnetic transition-metal monolayers, this relationship is
approximately satisfied when 
$\xi \approx \xi_{\rm SO}$ \cite{bruno89}. In our case
the $\xi$ is not directly related to $\xi_{\rm SO}$; it is smaller and
depends weakly on $J$. Notice however that $\xi(J)$ is positive and
Eq.~\ref{brunoeq} captures the connection between the anisotropic character of
the orbital moment and the sign of the anisotropy energy as a function
of $J$.

\section{Discussion}
We have undertaken a theoretical study of magnetic anisotropy 
in small magnetic nanoparticles 
that is motivated in part by an experimental study 
of 2 ML thick Co nanoplatelets
grown by M.H. Pan et Al \cite{kennanolett05}.  
Our study is based on a tight-binding model with
short range exchange and atomic SO-interactions. Experimentally, the clusters
are found to possess very high anisotropy energy per atom,
approximately one order of magnitude larger than bulk value, with
perpendicular easy directions (PED).

In addition to the experimental clusters, identified as having an FCC
structure truncated in parallel to the (111) crystal plane and a modified
lattice constant approximately equal to that of Si, we have studied the
clusters resulting from instead choosing the plane of truncation in parallel
to to (001) crystal plane. We found that the clusters with (111) geometry are more
prone to forming magnetic anisotropy energy landscapes with perpendicular easy
directions, an effect that can be traced to the difference in coordination
number between the (111) and (001) geometries.

In our study the model parameter that influences the shape of the anisotropy
landscape the most is found to be the intra-atomic $d$-electron exchange
coupling strength, $J$.  When $J$ is set close to a value which 
reproduce the experimentally observed mean magnetic moment, all truncation
schemes result in quasi-easy planes (QEP) for clusters larger than 50 atoms. For
clusters smaller than this, only the 2 ML (111) geometry is capable of producing a clear
bistable system, indicating a bias to PEDs for this symmetry. As $J$ is
increased to $1.5$ eV and $2.0$ eV all three geometries display PEDs, and the
anisotropy energy per atom tends to decrease as the exchange is increased
above the PED threshold value, identified as 1.5 eV. For $J=1.5$ eV and 2 ML
of (111) we find both PEDs and a very high MAE per atom, in
accordance with the experimental findings. 
Values of $J>1.5$ eV were considered here to investigate trends of the magnetic properties
in the asymptotic regime of large exchange coupling,   
albeit they are should not be regarded as physical.

Our simple model is able to produce the approximate scale of the MAE
in very anisotropically shaped nanoparticles, to provide a sense 
of its dependence on electronic structure details, and an indication of 
its complex dependence on band-filling, exchange interactions strength, and 
other electronic structure parameters.  
Moreover, the model shows that
the orbital magnetic moment is strongly anisotropic
with a magnitude greater than bulk, and its behavior as a function of the 
model parameters reflects the behavior of the MAE.
Nevertheless, when the 
phenomenological exchange constant of the model is chosen 
to reproduce the magnitude of 
the magnetization per atom, the anisotropy is underestimated and the 
sense of the anisotropy does not agree with the 
experiment of Ref.~\onlinecite{kennanolett05}.
Among the effects which we have 
neglected that could account for this discrepancy, 
two obvious possibilities are 
lattice-matching strains in the nanoparticle and hybridization between 
nanoparticle substrate.  In our calculations we have simply truncated the Co FCC
structure with the normal Co lattice constant and used bulk values for the tight-binding model hopping parameters.  
The Co nanoparticles that motivated this study are in all likelihood registered 
epitaxially to the substrate, stretching the inter-atom distances in the platelet plain 
and perhaps reducing the inter-atom distance between the two planes.  These 
altered atom-atom distances will alter the hopping parameters in a way that is likely
described at least approximately by bulk scaling properties\cite{harrison_es}. 
This additional anisotropy should likely be mainly uniaxial and could easily 
change the sense of the MAE. 
Hybridization with the substrate, corrugation of the
surface, and inter-cluster magnetic dipole interaction are all sources of 
MAE neglected in our model. However, experimental
results indicate that the effect of hybridization with Si is strongly
diminished by introducing the 0.5 ML Al spacer layer. In addition, it can be
seen that the effect of corrugation of the surface is minimal, by direct
inspection of the STM-extracted Co cluster height curve. 
As long as the
inter-cluster distance is sufficiently large (which it appears to be), we can
most likely neglect the inter-cluster magnetic dipole interaction.  

Our calculations suggest an alternative surface-physics based strategy for 
creating strongly anisotropic magnetic nanoparticles, namely 
the growth of Co clusters on a (001) Si surface covered with
0.5 ML of Al. To the best of our knowledge, such an attempt has not yet been
made. Recent results \cite{park_chae05} indicate that the 0.5 ML of Al will form
dimers oriented parallel to the dimer ridges formed by the Si. Of course,
the Si (001) surface is not homogeneous, but consists of domains of oriented
dimer ridges, highly dependent on the annealing process. By tuning the Al deposition rate, it should be possible to
obtain a substrate surface that is suitable for the growth of Co
squares or rectangles. If this Co structure turns out to form a 2 ML (001)
cluster, then that would result in a system with an
anisotropy energy 
higher than the one for nanoplatelets grown on a (111) surface.

\section{Acknowledgments}
We would like to thank Chih-Kang Shih
for interesting
discussions and Aleksander Cehovin for all his help with the computer
codes.
This work was supported in part by the Swedish Research Council
under Grant No:621-2001-2357 and by the Faculty of Natural Science
of Kalmar University.  AHM acknowledges support from the 
Welch Foundation and from the NSF under grant DMR-0115947.

\bibliography{./Biblio}

\end{document}